\documentclass{tlp}

\usepackage{pdfpages}

\usepackage{dsfont}
\usepackage{wrapfig} 
\usepackage{amsmath}
\usepackage{graphicx}
\usepackage{multirow}
\usepackage{relsize}
\usepackage{multicol}
\usepackage{multirow}
\usepackage[justification=centering]{caption}
\usepackage[list=true]{subcaption}
\usepackage{amssymb}
\usepackage{rotating}
\usepackage{comment}
\usepackage{pgfplots}
\usepackage{xspace}
\usepackage{booktabs}

\usepackage{xcolor}
\definecolor{linkcolor}{RGB}{52,59,144}
\usepackage[hidelinks, colorlinks=true, allcolors=linkcolor]{hyperref}
\usepackage{url}

\usepackage{ntheorem}

{
\theoremindent=.5cm
\theoremheaderfont{\kern-.5cm\normalfont\slshape}
\theorembodyfont{\normalfont}
\theoremseparator{.}
\theoremprework{\setlength\parindent{0cm}}
\newtheorem{example}{Example}
}

\usepackage{listings} 
\usepackage{textcomp}
\newcommand{\code}{\lstinline[style=MyInline]}

\lstnewenvironment{tablstcode}
{\lstset{style=myASP,numbers=none,xleftmargin=0cm,belowskip=-0.8 \baselineskip,aboveskip=-0.6 \baselineskip}}
{}
\lstdefinestyle{tree}
{
  basicstyle = \small\ttfamily\color{PrologPredicate},
  basewidth = 0.5em,
  moredelim = {[s][\color{PrologString}]{ \{}{\} }},
  moredelim = {*[s][{\color{PrologVar}}]{(}{)}},
  literate     =
  {.\\=.}{{\ \char"5C=\ }}3
  {\\=}{{\ \char"5C=\ }}3
  {.<.}{{\ \#<\ }}4
  {.>.}{{\ \#>\ }}4
  {=}{{\ =\ }}3
  {.=.}{{\ \#=\ }}3
  {.=<.}{{\ \#=<\ }}5
  {.>=.}{{\ \#>=\ }}5
}
\lstdefinestyle{MyInline}
{
  basicstyle = \relsize{-0.25}\ttfamily\color{PrologPredicate},
  breaklines = true,
  breakatwhitespace = true,
  upquote = true,
}
\lstdefinestyle{MyProlog}
{
  keywords = {},
  upquote = true,
  basicstyle = \ttfamily\color{PrologPredicate},
  basewidth = 0.48em,
  moredelim = {**[s][\color{PrologString!80!black}]{'}{'}},
  moredelim = {**[is][\color{PrologString!80!black}]{"}{"}},
  moredelim = {**[is][\color{PrologComment}]{`}{`}},
  moredelim = {**[is][\color{PrologPredicate}]{@}{@}},
  moredelim = {*[s][\color{PrologVar}]{(}{)}},
  moredelim = {*[s][\color{PrologOther}]{:-}{.}},
  moredelim = {*[s][\color{red}]{/*}{*/}},
  commentstyle = \mdseries\color{PrologComment},
  morecomment=[l]\%,
  morecomment=[s]{/*}{*/},
  literate     =
  {\\$}{{\$}}1
  {&(}{{\color{PrologOther}(}}1
  {&)}{{\color{PrologOther})}}1
  {&.}{{.}}0
  {\#}{{\ \#}}1
  {.=.}{{\,\#=\,}}3
  {.<.}{{\,\#<\,}}3
  {.>.}{{\,\#>\,}}3
  {.=<.}{{\,\#=<\,}}4
  {.>=.}{{\,\#>=\,}}4
  {=}{{\,=\,}}2
  {<}{{\,<\,}}2
  {>}{{\,>\,}}2
  {=<}{{\,=<\,}}3
  {>=}{{\,>=\,}}3
  {==}{{\,==\,}}3
  {.\\=.}{{\char"5C=}}3
  {\\=}{{\char"5C=}}3
  {,}{{\footnotesize,}}1
  {;}{{\footnotesize;}}1,
}
\lstdefinestyle{MyASP}
{
  keywords = {},
  upquote = true,
  basicstyle = \ttfamily\color{PrologPredicate},
  basewidth = 0.52em,
  moredelim = {*[s][\color{PrologOther}]{:-}{.}},
  moredelim = {*[s][\color{PrologOther}]{\{}{\}}},
  moredelim = {[s][\color{PrologVar}]{(}{)}},
  commentstyle = \color{PrologComment},
  morecomment=[l]\%,
  morecomment=[l]?,
  literate     =
  {..}{..}2
  {,}{{\footnotesize,}}1
}
\usepackage{xcolor}
\definecolor{PrologPredicate}{RGB}{0,0,200}
\definecolor{PrologVar}      {RGB}{145,032,039}
\definecolor{PrologComment}  {RGB}{169,082,044}
\definecolor{PrologOther}    {rgb}{0.2,0.2,0.2}
\definecolor{PrologString}   {RGB}{70,100,200}

\usepackage{morefloats}
\usepackage[
disable,
textwidth=0.13\textwidth,textsize=footnotesize]{todonotes}
\newcommand{\jahcommin}[1]{\todo[inline,backgroundcolor=green,linecolor=yellow]{ \textsf{ JAH: #1}}\xspace}

\newcommand{\mclcommin}[1]{\todo[inline,bordercolor=yellow,linecolor=yellow]{\textsf{ MCL: #1}}}

\newcommand{\myurl}[1]{\href{http://platon.etsii.urjc.es/~jarias/papers/spatial-iclp22/#1}{\nolinkurl{#1}}}
\newcommand{\redsection}{\vspace*{0em}}

\begin{document}

\lefttitle{Cambridge Author}

\jnlPage{1}{8}
\jnlDoiYr{2021}
\doival{10.1017/xxxxx}

\title[BIM using CLP]{Building Information Modeling Using\\ Constraint
  Logic Programming%
  \thanks{Work partially supported by EIT Digital, EU H2020 project
    BIM4EEB (Grant 820660), MICINN projects RTI2018-095390-B-C33
    InEDGEMobility (MCIU/AEI/FEDER, UE), PID2019-108528RB-C21 ProCode,
    Comunidad de Madrid project S2018/TCS-4339 BLOQUES-CM co-funded by
    EIE Funds of the European Union, US NSF (Grants IIS 1718945, IIS
    1910131, IIP 1916206), DoD, and Amazon.}}

\begin{authgrp}
\author{Joaqu{\'\i}n Arias$^1$ \hfill Seppo T\"orm\"a$^2$  \hfill Manuel Carro$^{3,4}$ \hfill Gopal Gupta$^5$} 
\affiliation{$^1$CETINIA, Universidad Rey Juan Carlos, Madrid, Spain \hspace{3em} $^2$ VisuaLynk Oy, Espoo, Finland\\
     $^3$Universidad Politécnica de Madrid, Spain \hspace{3em} $^4$IMDEA Software Institute, Pozuelo, Spain\\
   $^5$University of Texas at Dallas, Richardson, USA}
 
\end{authgrp}

\history{\sub{xx xx xxxx;} \rev{xx xx xxxx;} \acc{xx xx xxxx}}

\maketitle

\begin{abstract}

  Building Information Modeling (BIM) produces three-dimensional
  object-oriented models of buildings combining the geometrical
  information with a wide range of properties about materials,
  products, safety, to name just a few.  BIM is slowly but inevitably
  revolutionizing the architecture, engineering, and construction
  (AEC) industry.
  Buildings need to be compliant with regulations about stability,
  safety, and environmental impact. Manual compliance checking is
  tedious and error-prone, and amending flaws discovered only at
  construction time causes huge additional costs and delays.
  Several tools can check BIM models for conformance with
  rules/guidelines. For example, Singapore’s CORENET e-Submission
  System checks fire safety.
  But since the current BIM exchange format only contains basic
  information about building objects, a separate, ad-hoc model
  pre-processing is required to determine, e.g., evacuation
  routes.
  Moreover, they face difficulties in adapting existing built-in rules
  and/or adding new ones (to cater for building regulations, that can
  vary not only among countries but also among parts of the same
  city), if at all possible.

  We propose the use of logic-based executable formalisms (CLP and
  Constraint ASP) to couple BIM models with advanced knowledge
  representation and reasoning capabilities.  Previous experience
  shows that such formalisms can be used to uniformly capture and
  reason with knowledge (including ambiguity) in a large variety of
  domains.
  Additionally, incorporating checking within design tools makes it
  possible to ensure that models are rule-compliant at every step.
  This also prevents erroneous designs from having to be (partially)
  redone, which is also costly and burdensome.
  To validate our proposal, we implemented a preliminary reasoner
  under CLP(Q/R) and ASP with constraints and evaluated it with
  several BIM models.

  Under consideration for acceptance in Theory and Practice of Logic Programming
  (TPLP). \vspace{-0.5em}

\end{abstract}


\redsection
\section{Introduction}

Building Information Modeling
is a digital technology that is changing the Architecture,
Engineering, and Construction (AEC) industry.
It combines the three-dimensional geometry with non-geometrical
information of a building in an object-oriented model that can be
shared among actors over the construction lifecycle.
To facilitate the exchange of BIM models, \cite{IFC} has developed
Industry Foundation Classes (IFC), an open, vendor-neutral exchange
format.

Since 2016, the UK Government has required the Level 2 of BIM
maturity for any public construction project, where each discipline
generates specific models following the BIM standard. Once these
specific models are built, an Automated Code Compliance Checking tool,
e.g., BIM's Solibri Model Checker (SMC), provides basic architectural
checks, to verify the completeness of information and
detect the intersection of building components, among other things.
Additionally, automated tools check models in IFC format for
conformance with specifications, codes, and/or guidelines. For
example, the CORENET BIM e-Submission by \cite{Corenet} can be used to
check fire safety.
However, since the IFC format only represents basic building objects
and static information of their properties, pre-processing of the
model is required to, e.g., determine evacuation routes.
Moreover, these tools offer limited scope for customization or
flexibility and it is not easy to modify the implemented rules
and/or create new ones.

The domain of construction modeling
needs several capabilities: geometrical reasoning (including
arithmetical/mathematical capabilities and qualitative position
knowledge), reasoning about symbolic/conceptual knowledge, and
reasoning in the presence of vague concepts and/or incomplete
information (e.g., whether or not the outdoor space is safe depends on
details that are not yet known at this level of the design).
In addition, since part of the reasoning involves regulatory codes and
standards, a certain degree of ambiguity and
discretionary decisions are expected.

Interestingly, these different types of reasoning are not layered: a
model cannot be validated by first checking structural integrity
(i.e., that walls do not overlap or that columns are not placed where
a door is expected to be placed), then positional reasoning, and then
legal compliance. Legal requirements in this domain include
restrictions on sizes, areas, distances, relative positions, etc.
Therefore, a formalism suitable for checking (and, if executable, for
generating alternative models) has to be able to seamlessly capture
(and reason with) all of these types of information simultaneously.
Moreover, since regulations differ not only between countries, but
also among states/regions within a country, they must be easy to write
and, since they also change in time, to modify.

We believe that a formalism based on logic programming can meet many,
if not all, of the above requirements: a successful answer to a query
can determine that a model meets all the requirements.  Different
answers (or models) to a query may give alternative designs that
satisfy the requirements.
There exist query languages for BIM, such as BimSPARQL, by
\cite{zhang2018bimsparql}, and several logic-based proposals, e.g., by
\cite{pauwels2011a}, \cite{zhang2013a}, \cite{solihin2015simplified},
\cite{lee2016a}, and \cite{li2020non}, that validate our approach
because they show that minimal proof-of-concept tools have improved
reasoning capabilities w.r.t.\ commercial off-the-shelf BIM Sofware.
However, they all report limitations in the representation of
geometrical information and/or in the flexibility of the proposal to
adapt the code and/or the evaluation engine for different scenarios.

We propose to use tools integrating Constraint Logic Programming
with ASP to model dynamic information and restrictions in BIM models
and to enable the use of logic-based methodologies such as model
refinement. The main contributions of this paper are:
\begin{itemize}
\item A library, based on Constraint Answer Set Programming (CASP),
  that allows unified representation of geometrical and
  non-geometrical information.
\item The prototype of a preliminary 3D reasoner under Prolog with
  CLP(Q/R) that we evaluate with several BIM models.
\item The outline of an alternative implementation of this spatial
  reasoner under CASP, using s(CASP), by~\cite{scasp-iclp2018}, a
  goal-directed implementation of CASP.
\item Evidence for the benefits of using s(CASP) in BIM model
  evaluation: (i) it has the \emph{relevance} property, (ii) it can
  generate justifications for negative queries, and (iii) it makes
  representing and reasoning with ambiguities easier.

\end{itemize}

The ultimate goal of this work is to shift from BIM verification to
BIM refinement and to facilitate the implementation of new
specifications, construction standards, etc.

\redsection
\section{Background}

This section briefly describes (i) Building Information Modeling
(BIM), 
(ii) the industry foundation
classes (IFC), a standard for openBIM data exchange, and (iii)
s(CASP), a goal-directed implementation of constraint answer set
programming.

\redsection
\subsection{BIM + IFC}

Building information modeling (BIM)
has evolved from object-based parametric 3D modeling.
Combining geometrical information with other properties (costs,
materials, process, etc.)  makes it possible to have a range of new
functionalities, including cost estimations, quantity takeoffs, or
energy analysis.
The goal of BIM is to achieve a consistent and continuous use of
digital information throughout the entire life cycle of a facility,
including its design, construction, and operation.
BIM is based on a digital model and intends to raise productivity while
lowering error rates, as mistakes can be detected and resolved before
they become serious problems during construction and/or operation.
The most important advantages lie in the direct use of analysis and
simulation tools on these models and the seamless transfer of data
for the operation phase.
Today, there are numerous BIM authoring tools, such as Revit, ArchiCAD, Tekla Structures, or Allplan, that provide the basics for realizing BIM-based construction
projects.

\cite{IFC} has developed BIM interoperability technologies, the most important of which is IFC (Industry Foundation Classes), a common data model for
representing buildings.
IFC is standardized as ISO 16739 to improve BIM data interoperability
between heterogeneous BIM authoring tools and applications in their
disciplines.
The IFC schema is an extensive data model that logically encodes (i) the identity
and semantics (name, identifier, type), (ii) the characteristics or
attributes (material, color, thermal properties), and (iii) the
relationships (including locations, connections, and ownership) of (a)
objects (doors, beams), (b) abstract concepts (performance, costing),
(c) processes (installation, operations), and (d) people (owners,
designers, contractors, suppliers). For example, the IFC label
\emph{IfcBeam} is used to identify the beams (part of the structure of
a building that supports heavyweight).

IFC allows describing how a facility is designed, how it can be
constructed, and how its systems will function.  It defines 
building components, manufactured products, mechanical/electrical
systems, as well as more abstract models for structural analysis,
energy analysis, cost breakdowns, work schedules, etc.
IFC is in development since 1994 and now specifies close to one
thousand different entity types. IFC 4.0.1.2 was approved as ISO
standard 16739 in 2017. The specification of IFC5 is currently in
progress.

\redsection
\subsection{s(CASP)}

s(CASP), presented by \cite{scasp-iclp2018}, extends the
expressiveness of Answer Set Programming systems, based on the stable
model semantics by \cite{gelfond88:stable_models}, by including
predicates, constraints among non-ground variables, uninterpreted
functions, and, most importantly, a top-down, query-driven execution
strategy.
These features make it possible to return answers with non-ground
variables (possibly including constraints among them) and to compute
partial models by returning only the fragment of a stable model that
is necessary to support the answer to a given query.

In s(CASP), thanks to the constructive negation,
\code|not p(X)| can return bindings for \code{X} 
for which the call \code{p(X)} would have failed.
Additionally, thanks to the interface of s(CASP) with constraint
solvers, sound non-monotonic reasoning with constraints is possible.
s(CASP), like other ASP implementations and unlike Prolog, handles
non-stratified negation.

\begin{example}\label{exa:size}
  Consider the program \code{size(r1,S):- S#>=21} (see \myurl{size.pl}),
  For the query \code{?- not size(r1,S)}, s(CASP) returns the
  model %
  \code|{not size(r1,S $|$ {S #< 21})}|.
\end{example}

\begin{example}\label{exa:bigsmall} The following program, in
  \myurl{kitchen.pl}, models that the room \code{r1} is either small
  or big and it is a kitchen:
\begin{lstlisting}[style=MyProlog]
small(r1) :- not big(r1).
big(r1) :- not small(r1).
kitchen(r1).
\end{lstlisting}

\noindent
The top-down evaluation of the non-stratified negation in lines 1-2
detects a loop having an even number of intervening negations (and
\emph{even loop}).  When this is discovered,
the truth/falsehood of the atoms involved is guessed to generate
different models whose consistency  is subsequently checked.
In this example, there are two possible models, and given a query it
returns the \emph{relevant} partial model (if it exists):
\begin{itemize}
\item[] \code{?- small(r1).} \hfill returns %
  \code|{small(r1), not big(r1)}|
\item[] \code{?- big(r1).} \hfill returns %
  \code|{big(r1), not small(r1)}|
\item[] \code{?- kitchen(r1).} \hfill returns \code|{kitchen(r1)}|
\item[] \code{?- big(r1), small(r1).} \hfill returns
  \code{no models}
\end{itemize}
\end{example}

In addition to default negation, s(CASP) supports classical negation,
marked with the prefix~'\code{-}', to capture the explicit evidence
that a literal is false: \code{not small(r1)} expresses that we have
no evidence that \code{r1} is small (we can not prove it), and
\code{-small(r1)} means that we have \emph{explicit} evidence (there
is proof) that \code{r1} is not small.

Finally, s(CASP) provides a mechanism to present justifications in
natural language.
Both plain text and user-friendly, expandable HTML can be generated
(e.g., \myurl{small_r1.txt} and \myurl{small_r1.html} show the
justification for the query %
\code{?- small(r1)} in Example~\ref{exa:bigsmall}).

\redsection
\section{Modeling vague concepts}
\label{sec:model-vague-conc}

We present now a proposal to represent vague concepts
using s(CASP).
The formal representation of legal norms to automate reasoning and/or
check compliance is well known in the literature.  There are several
proposals for deterministic rules.  However, none of the existing
proposals, based on Prolog or standard ASP, can efficiently represent
vague concepts due to unknown information, ambiguity, and/or
administrative discretion.

\begin{example}\label{exa:rule}
  Considering the following norm from a building regulation:

  \begin{quote}
    In the room there is at least one window, and each window must be
    wider than 0.60 m. If the room is small, it can be between 0.50
    and 0.60 m wide.
  \end{quote}

  We can encode this norm using default negation:

\begin{lstlisting}[style=MyProlog]
requirement_a(Room):- window_belongs(Window,Room), width(Window,Width), 
                      Width#>$0.60$, not small(Room).
requirement_a(Room):- window_belongs(Window,Room), width(Window,Width), 
                      Width#>$0.50$, small(Room).
\end{lstlisting}

  \noindent
  However, without information on the size of the room or what is the
  criteria to consider that a room is small, 
  it is not possible to determine whether the room is small and only
  the first rule would succeed.
\end{example}

To encode the absence of information we propose the use of the stable
model semantics by \cite{gelfond88:stable_models}, which makes it
possible to reason about various scenarios
simultaneously, e.g., in the previous example we can consider two
scenarios (models): in one a given room is small and in the other, it
is not.

\begin{figure}
\begin{lstlisting}[style=MyProlog, basewidth=.46em]
room(r1).      room(r2).      room(r3).      room(r4).
room(r5).      room(r6).      room(r7).      room(r8).
size(r1, 25).        size(r2, 5).        size(r3, 15).

evidence(Room, small) :- size(Room,Size), Size#<10.
-evidence(Room, small) :- size(Room,Size), Size#>20.

small(Room) :- evidence(Room,small).
-small(Room) :- -evidence(Room,small).
small(Room) :- not evidence(Room,small), not -small(Room).               
-small(Room) :- not -evidence(Room,small), not small(Room).

room_is(Room,big) :- room(Room), -small(Room).
room_is(Room,small) :- room(Room), small(Room).
\end{lstlisting}
  
  \caption{Code representing vague and unknown information (available
    at \myurl{room.pl}).}
  \label{fig:room}
\end{figure}

\begin{example}[Example~\ref{exa:rule} (cont.)]
  \label{exa:rule2}
  Fig.~\ref{fig:room} models a hotel with eight rooms,
  for which we only know the size of three (lines 1-3).
  Following the patterns by~\cite{ariaslaw}, lines 8-9 make
  it possible to reason considering (i) unknown information (size of
  rooms $r4$ to $r8$), and/or (ii) vague concepts: line 5 states that
  a room smaller than 10 m$^2$ is small and line 6 states that a room
  larger than 20 m$^2$ is not small.  However, it is not clear
  whether rooms with size between 10 and 20 m$^2$ are small or not.
  Line 8 captures that there is evidence that the room is small, line
  9 captures the case when there is a explicit evidence (there is
  proof) that the room is not small, and lines 10-11 generate two
  possible models otherwise.
  Finally, \code{room_is/2} determines whether a room is small or not based
  on evidence and/or assumptions: for room $r1$ the query
  \code{?-room_is(r1,Size)} returns \code{S=big}, for $r2$ it
  returns \code{S=small}, and for the other rooms it returns both
  alternatives: \code{S=small} assuming that \code{small(r3)} holds
  and \code{S=big} assuming that %
  \code{-small(r3)} holds.
\end{example}

\redsection
\section{Modeling and manipulating 3D objects}

We will now describe our proposal to model and manipulate geometrical
information used to represent 3D objects which will be used to model
buildings, infrastructures, etc.\ using constraints.
Additionally, we show how the operations that manipulate these objects
can also be used to infer new knowledge and/or to verify that
geometrical data and non-geometrical information are consistent at
each stage of the project development.

\redsection
\subsection{Representing 3D objects using linear equations}

Any 3D object can be approximated as the union of convex shapes.
The simplest shape to represent
with linear equations is a box with edges parallel to the axes of
coordinates.  Assuming that $P_a$ and $P_b$ are opposing vertices with
$P_a$ being the one closest to the coordinate origin, the box is a set
of points $P$ with coordinates $(X, Y, Z)$  (resp.\ $(X_a, Y_a, Z_a)$
for $P_a$ and similarly for $P_b$) 
such that 
\begin{displaymath}
X \geq X_a \land X < X_b \land Y \geq Y_a \land Y < Y_b
\land Z \geq Z_a \land Z < Z_b 
\end{displaymath}

Equations of this type can be easily managed using a simple linear
constraint solver.  In this paper, we use the well-known CLP(Q) solver
by \cite{holzbaur-clpqr}\footnote{We use a constraint solver
  over rationals to avoid the loss of precision associated with the
  use of reals.} and we represent such an object with the term
\code{point(X, Y, Z)} where \code{X}, \code{Y}, and \code{Z} are
variables adequately constrained as shown before.
When a complex object has to be decomposed into convex shapes $S_i$,
we define this object as the set of points $P$ such that
$P \in S_1 \lor P \in S_2 \lor \dots \lor P \in S_n$.
Since CLP(Q) does not provide logical disjunction as part of the
constraint solver's operations, we explicitly represent the union of objects as a
list of convex shapes %
\code{[convex(Vars_1), convex(Vars_2), $\dots$]}, where \code{Vars_x}
encodes the variables that carry the constraints corresponding to the
linear equations of each $S_i$.

\begin{example}
  A 3D box whose defining vertices are $P_a$ and $P_b$ (represented,
  resp.\ as \code{point(Xa,Ya,Za)} and \code{point(Xb,Yb,Zb)}), is
  encoded as:

\begin{lstlisting}[style=MyProlog]
box(point(Xa,Ya,Za), point(Xb,Yb,Zb), [convex([X,Y,Z])]) :-
    X#>=Xa, X#<Xb, Y#>=Ya, Y#<Yb, Z#>=Za, Z#<Zb.
\end{lstlisting}

  \noindent
  where the box is represented with a list that contains a single
  convex shape.
\end{example}

Most objects in the definition of a building are extruded 2D convex
polygons,
and therefore having an explicit operation for this case is
useful and it illustrates the power of using CLP for modeling
building structures.

\begin{example}
  Given a 3D object defined by its base (a convex polygon determined by its 
  vertices $A,B,C, \dots$ in clockwise order) and its height $H$,
  its representation is:

\begin{lstlisting}[style=MyProlog]
poly_extrude(Vertices, H, [convex([X,Y,Z])]) :-
    Vertices = [point(_,_,Za), _, _ | _],     % At least three points
    polygon(Vertices,[X,Y]), Z#>=Za, Z#<Za+H.
polygon([_], _).
polygon([point(Xa,Ya,_), point(Xb,Yb,_) | Vs], [X,Y]) :-
     (Xb-Xa) * (Y-Ya) - (X-Xa) * (Yb-Ya) #=< 0,
     polygon([point(Xb,Yb,_) | Vs], [X,Y]).
\end{lstlisting}
\end{example}

\vspace{-1em}
\redsection
\subsection{Operations on $n$-dimensional objects under CLP(Q)}
\label{sec:oper-n-dimens}

\begin{figure}
  \centering
  \begin{lstlisting}[style=MyProlog, numbers=left]
% Union = ShA $\cup$ ShB 
shape_union(ShA, ShB, Union) :- append(ShA, ShB, Union).

% Intersection = ShA $\cap$ ShB 
shape_intersect(ShA, ShB, Intersection) :-
    shape_intersect_(ShA, ShB, [], Intersection).
shape_intersect_([],_,ShInt,ShInt) :- !.
shape_intersect_(_,[],ShInt,ShInt) :- !.
shape_intersect_([Sh1|Sh1s],[Sh2|Sh2s],ShInt0,ShInt) :-
    convex_intersect(Sh1, Sh2, Sh12),
    shape_union(ShInt0,Sh12,ShInt1),
    shape_intersect_([Sh1], Sh2s, ShInt1,ShInt2),
    shape_intersect_(Sh1s,[Sh2|Sh2s],ShInt2,ShInt).
convex_intersect(convex(Vars1),convex(Vars2),ShInt) :-
    &(   copy_term(Vars1,VarInt), copy_term(Vars2,VarInt), true ->
        ShInt = [convex(VarInt)]
    ;   ShInt = []                &).

% Complement = $\lnot\;$ ShA
shape_complement([], [convex(_)]).
shape_complement([convex(Vars)], NotSh) :-
    findall(convex(DualVars), dual_vars_clp(Vars, DualVars), NotSh).
dual_vars_clp(Vars, TempVars) :-
    dump_clp_constraints(Vars,TempVars,TempGeo), TempGeo \= [],      
    dual_clp(TempGeo,Dual), apply_clp_constraints(Dual).

% Subtract = ShA $\cap$ $\lnot\;$ ShB
shape_subtract([],_,[]) :- !.
shape_subtract(ShA,[],ShA) :- !.
shape_subtract([Sh1|Sh1s],ShB, Sh1Remains) :-
    convex_subtract(Sh1,ShB,Remain0),
    shape_subtract(Sh1s,ShB,Remain1),
    shape_union(Remain0,Remain1,Sh1Remains).
convex_subtract(Sh1,[Sh2|Sh2s],Sh1Remains) :-
    &(   convex_intersect(Sh1,Sh2,ShInt), ShInt == [] ->
        shape_subtract([Sh1],Sh2s,Sh1Remains)
    ;   shape_complement([Sh2], NotSh2),
        shape_intersect([Sh1], NotSh2, ShRemain0),
        shape_subtract(ShRemain0,Sh2s,Sh1Remains)    &).
\end{lstlisting}
\vspace{-.5em}
  \caption{Operations on an $n$-dimensional space using CLP(Q)
    \myurl{spatial_clpq.pl}.}
  \label{fig:operations}

  \vspace*{1em}

\begin{lstlisting}[style=MyProlog, numbers=left, basewidth=.43em]
% Union = ShA $\cup$ ShB 
shape_union(IdA, IdB, convex([X,Y])) :-  convex(IdA, X, Y).
shape_union(IdA, IdB, convex([X,Y])) :-  convex(IdB, X, Y).
% Intersection = ShA $\cap$ ShB 
shape_intersect(IdA, IdB, convex([X,Y])) :- convex(IdA, X, Y), convex(IdB, X, Y).
% Complement = $\lnot\;$ ShA
shape_complement(IdA, convex([X,Y])) :- not convex(IdA, X, Y).
% Subtract = ShA $\cap$ $\lnot\;$ ShB
shape_subtract(IdA, IdB, convex([X,Y])) :- convex(IdA, X, Y), not convex(IdB, X, Y).
\end{lstlisting}
\vspace{-.5em}
  \caption{Operations on a 2-dimensional space using s(CASP)
    \myurl{spatial_scasp.pl}}
  \label{fig:operations-scasp}

\end{figure}

In this section, we explain some basic operations (union, intersection,
complement, and subtraction) to manipulate the 3D objects that
describe the BIM project.
Fig.~\ref{fig:operations} shows a preliminary interface implemented
using CLP(Q) with the following four predicates that can be used to
manipulate shapes in any $n$-dimensional space:\footnote{We assume
  that union, intersection, etc.\ of shapes is understood as the
  corresponding operations on the sets of points contained inside the
  shapes.}

\begin{itemize}
\item \code{shape_union(ShA, ShB, Union)}: Given two shapes $ShA$ and
  $ShB$, it creates a new shape $Union$ that is the union, i.e.,
  $Union = ShA \cup ShB$.  Since every shape is, in general, the union
  of simpler convex shapes (represented as a list thereof), $Union$
  can simply be represented as a list containing the convex shapes of
  $ShA$ and $ShB$ (line 2).  Simplification (to, e.g., remove
  shapes contained inside other shapes) can be done, but we have not
  shown it here for simplicity.
  
\item \code{shape_intersect(ShA, ShB, Intersection)}: Given two
  shapes $ShA$ and $ShB$, it creates a new shape $Intersection$ which
  is the intersection, i.e., $Intersection = ShA \cap ShB$ (lines
  5-6).
  This is computed with the union of the pairwise intersections of
  the shapes in $ShA$ and $ShB$.
  Obtaining the intersection of two convex shapes is done by the
  CLP(Q) constraint solver, which determines
  the set of points that are in both shapes as those which satisfy the
  constraints of both shapes. In line 15, \code{copy_term/2}
  preserves the variables by generating a new set,
  \code{VarInt}.\footnote{In line 15, \code{true} is needed due to a
    well-known, subtle bug in the implementation of attributed
    variables of the underlying Prolog infrastructure used in this
    paper.}

\item \code{shape_complement(ShA, Complement)}: Given a shape $ShA$,
  it creates a new shape $Complement$ that contains every point in the
  $n$-dimensional space that is not in the shape $ShA$ (lines
  20-25). This is computed as the dual of $ShA$. From a logical point
  of view, it corresponds to the negation, so it is
  denoted as $Complement = \lnot\ ShA$.

\item \code{shape_subtract(ShA, ShB, Subtraction)}: Given two
  shapes $ShA$ and $ShB$, the subtraction is a new shape $Subtraction$
  that contains all points of $ShA$ that are not in $ShB$, i.e.,
  $Subtraction = ShA \cap \lnot\ ShB$ (lines 28-39).
  To compute the subtraction of a convex shape from $ShA$, we iteratively
  narrow its $n$-dimensional space, $C_0$, by selecting one convex shape from $ShB$,
  $Sh_i$, and computing the intersection of $C_{i-1}$ and the complement of
  $Sh_i$, i.e., $C_i = C_{i-1} \cap \neg Sh_i$. The 
  execution finishes when all shapes have been selected or
  the $i^{th}$ shape covers the remaining space $C_{i-1}$.

\end{itemize}

\begin{example}\label{exa:clpq} 
  These operations can be used with $n$-dimensional shapes.  For
  simplicity, let us consider 2D rectangles: \emph{r1} in yellow and
  \emph{r2} in blue):

\begin{lstlisting}[style=MyProlog]
obj(r1, [convex([X,Y])]) :- X#>=1, X#<4, Y#>=2, Y#<5.
obj(r2, [convex([X,Y])]) :- X#>=3, X#<5, Y#>=1, Y#<4.
\end{lstlisting}

  \vspace*{.5em}
  
  \begin{minipage}{.63\linewidth}

\begin{itemize}[wide]
\item The query
  \code{?- obj(r1,Sh1), obj(r2,Sh2),}
  
  \hfill \code{shape_intersect(Sh1, Sh2, Intersection)}

  returns:
  \code{Intersection = [convex([A,B])], }

  \hfill \code{A#>=3, A#<4, B#>=2, B#<4}
  
\item The query
  \code{?- obj(r1,Sh1), obj(r2,Sh2),}

  \hfill \code{shape_subtract(Sh1, Sh2, Subtraction)}

  returns:

  \code{Subtraction = [convex([A,B]),convex([C,D])],}
  
  \code{A#>=1,A#<3,B#>=2,B#<5, C#>=3,C#<4,D#>=4,D#<5}
\end{itemize}
  \end{minipage}
  \begin{minipage}{.35\linewidth}
    \raggedleft
    \includegraphics[width=.9\linewidth]{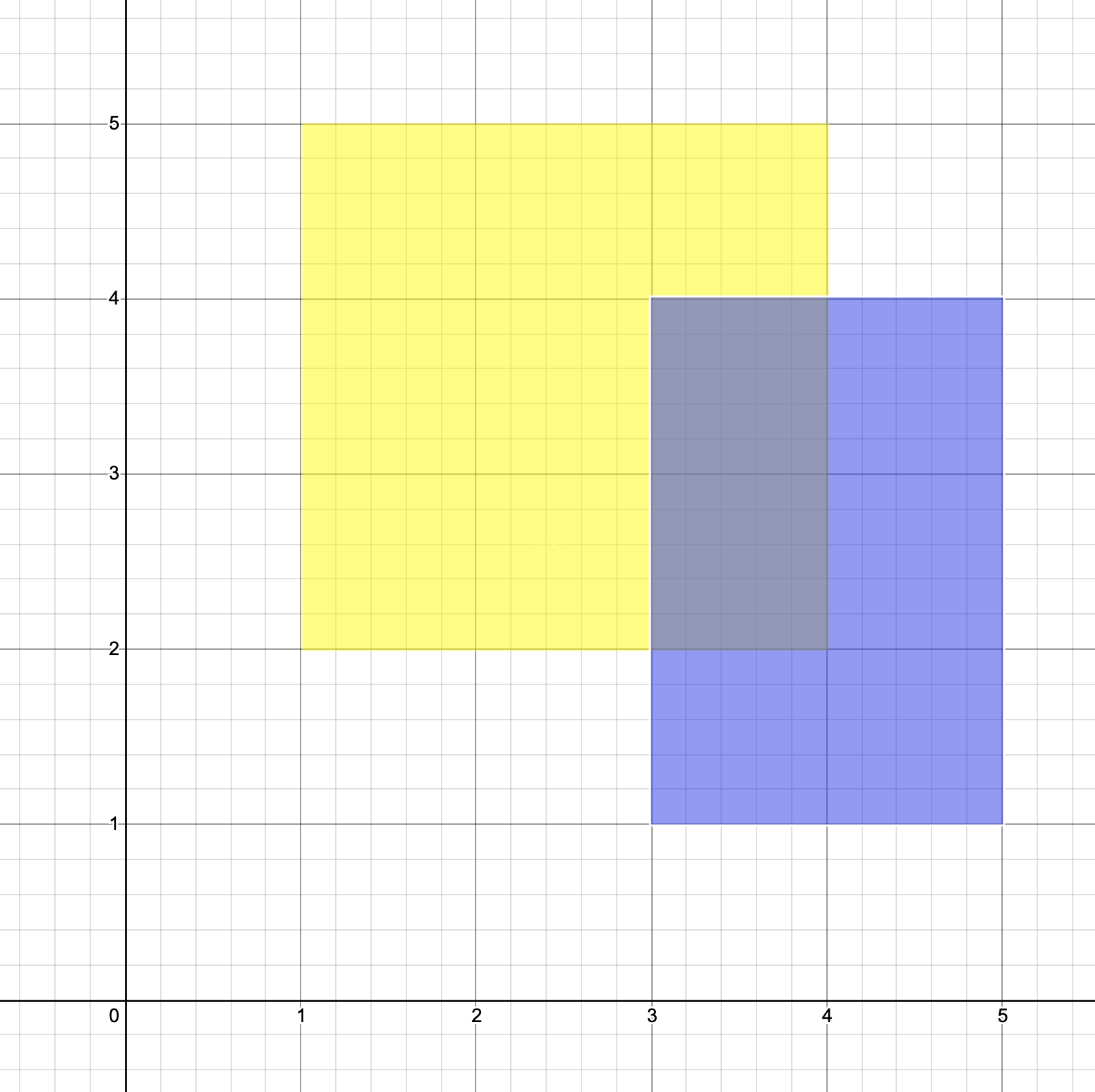}
  \end{minipage}
\end{example}

  \vspace*{.5em}

\begin{example}
  In addition, they can be combined to verify IFC properties and/or
  (non-)geometrical information contained in the BIM model.
  The predicate \code{window_belongs(W,R)}, in Example~\ref{exa:rule},
  can be defined from the geometry of \code{W} and \code{R}, i.e.,
  \code{W} belongs to \code{R} if the intersection returns a non-empty
  shape.
\end{example}

\redsection
\subsection{Operations on $\mathbf{n}$-dimensional shapes under  s(CASP)}
\label{sec:oper-n-dimens-1}

We will now sketch how the main operations on $n$-dimensional objects
can be defined using s(CASP).  We will take advantage of its ability
to execute ASP programs featuring variables with dense, unbounded
domains.
As a proof of concept, Fig.~\ref{fig:operations-scasp} shows the
encoding of the operations for 2-dimensional shapes, with slight
differences in the representation of the objects and shapes w.r.t.\
the representation used under Prolog:

\begin{itemize}

\item The predicates for union, intersection, complement, and
  subtraction receive the identifiers of the object(s), instead of the
  list of convex shapes.
  
\item A convex shape in $n$ dimensions is an atom with $n+1$ arguments.  Its first argument is the object identifier
  and the rest of the arguments are the variables used to define the
  convex shape, e.g., a 2D shape is an atom of the form
  \code{convex(Id,X,Y)}.

\item The representation of the convex shapes is part of the program,
  e.g., the rectangles $r1$ and $r2$ in Example~\ref{exa:clpq} are
  defined with the clauses:

\begin{lstlisting}[style=MyProlog]
convex(r1, X, Y) :- X#>=1, X#<4, Y#>=2, Y#<5.
convex(r2, X, Y) :- X#>=3, X#<5, Y#>=1, Y#<4.
\end{lstlisting}

\end{itemize}

This representation delegates the shape representation to be handled
as part of the constraint store of the program.  Therefore, a
non-convex object is represented with several clauses, one for each
convex shape, and a set of convex shapes is a set of answers obtained
via backtracking.

\begin{example}[Cont. Example~\ref{exa:clpq}] Let us consider the queries
  used in Example~\ref{exa:clpq} under s(CASP) with the encoding in
  Fig.~\ref{fig:operations-scasp}.

  \begin{itemize}[wide]
  \item The query
  \code{?-shape_intersect(r1,r2,Intersection)}
  returns:

  \code|Intersection = convex([A $|$ { A#>=3, A#<4 }, B $|$ { B#>=2, B#<4 }])|

\item The query
  \code{?-shape_subtract(r1,r2,Subtraction)}
  returns two answers:
  \code|Subtraction = convex([A $|$ { A#>=1, A#<3 }, B $|$ { B#>=2, B#<5 }]) ? ;|
  \code|Subtraction = convex([A $|$ { A#>=3, A#<4 }, B $|$ { B#>=4, B#<5 }])|
  
  \end{itemize}
   
\end{example}

\redsection
\section{Tracing (non)-monotonic changes in BIM models}
\label{sec:trace-non-monotonic}

Let us adapt the example presented by~\citep{scasp-iclp2018} in
Section 4.1, where different data sources may provide inconsistent
data, and a stream reasoner, based on s(CASP), decides whether the
data is valid or not depending on how reliable are the sources.

Here, instead of streams, we consider models.  A shared model is
updated by different experts, and updated models have to be merged to
generate the next model in the chain.

\begin{example}\label{exa:trace}
  Consider a shared BIM model that contains information about room
  ventilation, a heating boiler feed system, and a fire safety
  regulation that states:
  \begin{itemize}[wide=0.5em, leftmargin =*, nosep]
  \item If a gas boiler is used, the ventilation must be
    natural.\footnote{The ventilation of a room is natural if the area
      of its windows is at least 10 percent of its floor's.}
  \item If an electric boiler is used, the ventilation could be
    either natural or mechanical.
  \end{itemize}

  Initially, the shared BIM model has no ventilation or boiler
  restrictions.
  Later on, the architect modifies the model by reducing the size of
  the window in such a way that ventilation cannot be considered
  natural any longer due to the new size. To comply with the fire
  safety regulation (and maintain the consistency of the model) the
  architect selects an electric heating boiler.
  Simultaneously, the engineer modifies the model by selecting a gas
  boiler because it is more efficient than an electric boiler. This
  would force ventilation to be natural.

  Finally, when attempting to merge the updated models, an
  inconsistency is detected and the integration fails.
  A naive approach would broadcast the inconsistency to the architect
  and engineer, but we propose using a continuous integration reasoner
  to determine who is the expert whose opinion prevails and make a
  decision based on that.
  The other party needs then to be notified to confirm the
  adjustments.
\end{example}

\begin{figure}
  \begin{minipage}{.9\linewidth}
\begin{multicols}{2}
\begin{lstlisting}[style=MyProlog]
%% BIM Continuous Integration
valid_data(P,Data) :- 
     data(P,Data), 
     not canceled(P, Data).

canceled(P, Data) :- 
     higher_confidence(P1, P), 
     data(P1, Data1), 
     inconsistent(Data, Data1).
higher_confidence(PHi, PLo) :- 
     PHi.>.PLo.
%% Example
inconsistent(boiler(gas),
              ventilation(artificial)).
inconsistent(ventilation(artificial),
              boiler(gas)).
data(1,ventilation(X)).   
data(2,ventilation(natural)).
data(2,boiler(gas)).   
% data(3,ventilation(artificial)).
% data(3,boiler(electrical)).
% data(4,boiler(gas)).  
\end{lstlisting}    
  \end{multicols}
\end{minipage}
\caption{Code of the BIM continuous integration with an example.}
\label{fig:ci_code}
\end{figure}

Fig.~\ref{fig:ci_code} sketches a continuous integration reasoner,
adapted from the paper by \cite{scasp-iclp2018} and the encoding of
Example~\ref{exa:trace} (\myurl{Bim_CI.pl}). Its goal is the detection
of inconsistency in pieces of information provided by the different
stakeholders. And then, depending on a given criterion (e.g., their
responsibility/expertise), it would determine which data is valid (and
eventually who should amend that inconsistency).
Data labels are represented as
\code{data(Priority, Data)}, where \code{Priority} tells us the degree
of confidence in \code{Data};
\code{higher_confidence(PHi,PLo)} hides how priorities are encoded in
the data (in this case, the higher the priority, the more level of
confidence);
and \code{inconsistent/2} determines, in lines 13-16, which data items
are inconsistent (in this case, \code{ventilation(artificial)} and
\code{boiler(gas)}).
Lines 1-11, alone, define the reasoner rules: \code{valid_data/2}
states that a data label is valid if it is \emph{not canceled} by
another data label with more confidence.\footnote{Inconsistent data
  with the same confidence remain valid unless there is a more
  confident inconsistency.}
In this encoding the confidence relationship uses constraints, that
instead of being checked afterward prune the search, but it is
possible to define more complex rules, i.e., to determine who is more
expert/confident depending on the data itself (e.g., for discrepancies
in the dimensions of a beam, the structural engineer is the expert).

Lines 17-19 define that initially, \code{ventilation(X)} holds for
\textbf{all} \code{X}, but when the engineer selects
\code{ventilation(natural)} and \code{boiler(gas)} this data has more
confidence, so the query %
\code{?-valid_data(Pr,Data)} returns:
\code|{Pr=1, Data=ventilation(A), A\=artificial}| \mbox{because}
\code{boiler(gas)} is more reliable than \code{ventilation(X)},
\code|{Pr=2, Data=ventilation(natural)}|, and %
\code|{Pr=2, Data=boiler(gas)}|.
If we consider that the architect selection has more confidence than
the engineer's (by adding lines 20-21), the query
\code{?-valid_data(Pr,Data)}, returns:
\code|{Pr=1, Data=ventilation(A), A\=artificial}|, %
\code|{Pr=2, Data=ventilation(natural)}|, %
\code|{Pr=3, Data=ventilation(artificial)}|, and %
\code|{Pr=3, Data=boiler(electrical)}|. Note that now the answer
\code|{Pr=2, Data=boiler(gas)}| is not valid. %
Finally, by adding \code{data(4,boiler(gas))} (line 22), we observe
that answer \code|{Pr=3, Data=ventilation(artificial)}| is not valid.
As we mentioned before, s(CASP) also provides justification trees for
each answer (\myurl{Bim_CI.txt}) to support the inferred conclusions
so the user can check and/or validate the correctness of the final
results.

\redsection
\section{Evaluation}

The reasoner and benchmarks used in this paper are available at
\url{https://gitlab.software.imdea.org/joaquin.arias/spatial}, and/or
at
\url{http://platon.etsii.urjc.es/~jarias/papers/spatial-iclp22}. They
were run on a macOS 11.6.4 laptop with an Intel Core i7 at 2,6
GHz. under Ciao Prolog version 1.19-480-gaa9242f238
(\url{https://ciao-lang.org/}) and/or under s(CASP) version 0.21.10.09
(\url{https://gitlab.software.imdea.org/ciao-lang/scasp}).

A direct performance comparison of our prototype with implementations
in other tools may not be meaningful because they do not support the
representation of vague concepts and/or continuous quantities.
ASP4BIM by \cite{li2020non} overcomes most of the limitations of
previous tools (see Section~\ref{sec:related-work}) but, since it is
built on top of \emph{clingo}, it inherits limitations already pointed
out by \cite{arias-ec2022}.

Firstly, let us use the program \myurl{room.pl} (Fig.~\ref{fig:room}).
As we mentioned in Section~\ref{sec:model-vague-conc}, it returns
independent answers under s(CASP), i.e., for the query %
\code{?- room_is(Room,Size)} we obtain a total of \textbf{14 partial
  models}: one for room $r1$, another for room $r2$, and two for each
of other six rooms. On the other hand, the same program under clingo
(\myurl{room.clingo}) generates \textbf{64 models}: all possible
combinations such as \code{room_is(r1,big)} and
\code{room_is(r2,small)} appear in all of them and for each room $rX$,
32 models contains \code{room_is(rX, big)}, while the other 32 models
contains \code{room_is(rX, small)}. The exponential explosion in the
number of models generated by clingo reduces the comprehensibility of
the results (for 16 rooms it generates \textbf{16384 models}, while
s(CASP) generates only \textbf{30 models}).
Moreover, the goal-directed evaluation of s(CASP) not only makes it
possible to reason about specific rooms, but it also generates the
corresponding justification, e.g., \myurl{room_r1.txt} and
\myurl{room_r1.html} for room $r1$.

\begin{figure}
  \centering
  \begin{subfigure}{.36\linewidth}
    \includegraphics[width=1.1\linewidth]{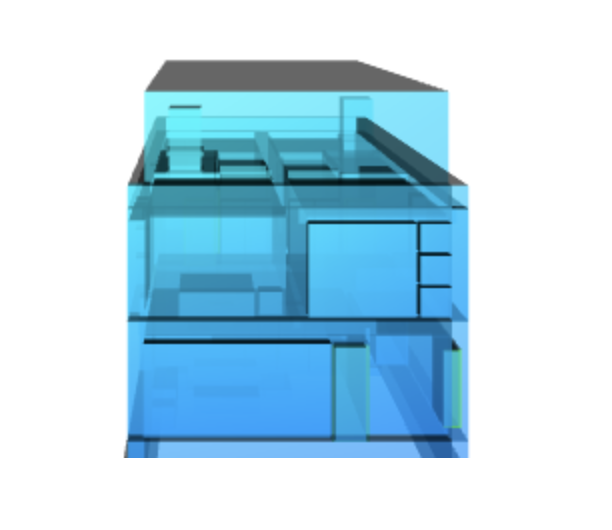}
  \vspace*{-1.5em}
  \caption{\myurl{Duplex_Q1.html}}
  \label{fig:x3d_a}
  \end{subfigure}
  \begin{subfigure}{.36\linewidth}
    \includegraphics[width=1.1\linewidth]{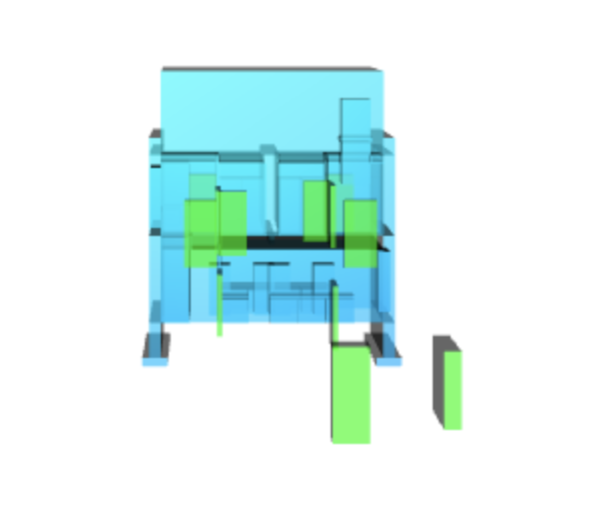}    
  \vspace*{-1.5em}
  \caption{\myurl{Duplex_Q2.html}}
  \label{fig:x3d_b}
\end{subfigure}
  \begin{subfigure}{.26\linewidth}
    \includegraphics[width=1.1\linewidth]{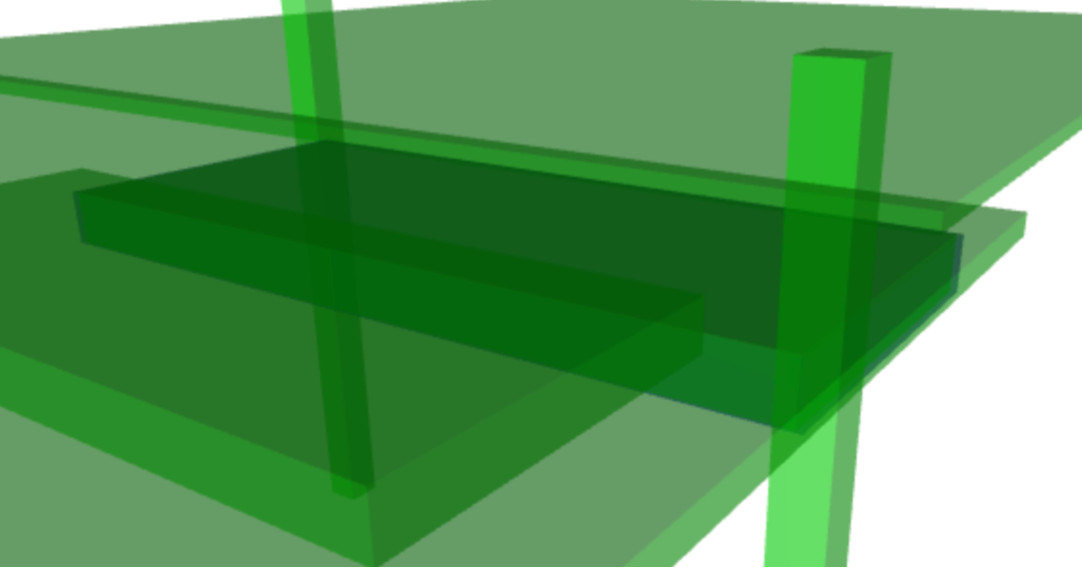}    
  \vspace*{-1em}
  \caption{\myurl{Office_Q1.html}}
  \label{fig:x3d_c}

    \includegraphics[width=1.1\linewidth]{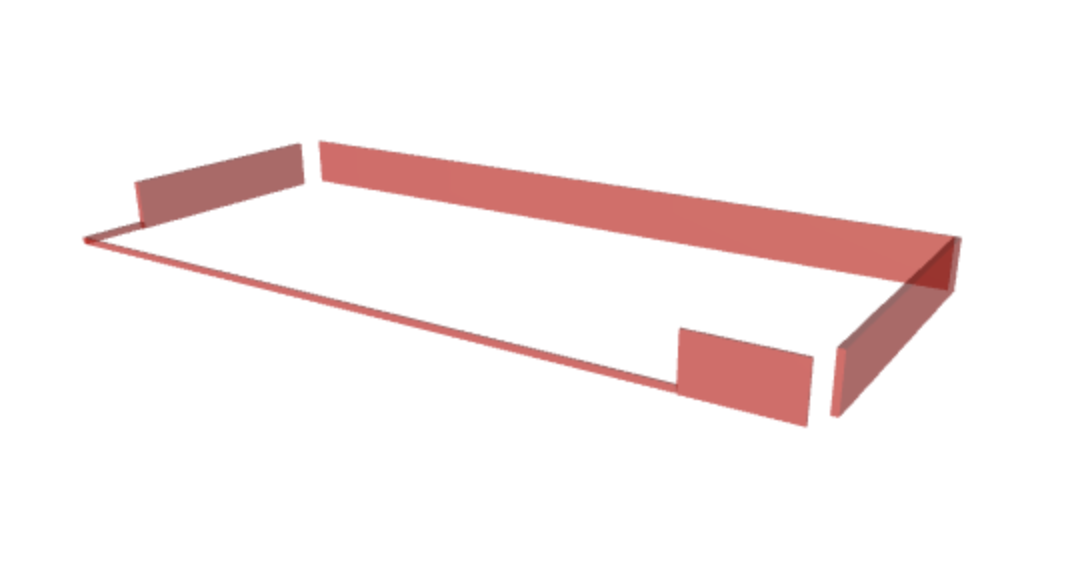}    
  \vspace*{-2em}
  \caption{\myurl{Office_Q2.html}}
  \label{fig:x3d_d}
  \end{subfigure}
  \caption{Images in x3d corresponding to the Duplex and Office BIM
    models.}
\end{figure}

Secondly, to validate the benefits of our proposal dealing with
geometric information, we have implemented a spatial reasoner, in
collaboration with VisuaLynk Oy, based on the spatial
interface described in Fig.~\ref{fig:operations}.
This spatial reasoner includes a graphic interface that translates 
the constraints back into geometry and generates 3D images with the results
for the queries using x3d.\footnote{Reference textbook for learning
  Extensible 3D (X3D) Graphics available at
  \url{https://bit.ly/3O1MqcH}}
%
%
The benchmarks used are: (i) the ERDC: \textbf{Duplex} Apartment Model
ERDC D-001
produced in Weimar, Germany for a design competition,
and (ii) the Trapelo St.\ \textbf{Office} (IFC4 Edition),
a 3-story office building where Revit HQ in Waltham is, which consists
of three models (Architecture, MEP, and Structure).

For the evaluation, we translated the IFC files of the models to
convert the geometrical data (and IFC labels) of the 286 objects of
the Duplex and approximately 5000 objects of the Office Building (3639
objects in the architecture model and 1322 objects in the structure
model) into Prolog facts.  We defined a predicate \code{object/5},
where the first argument is the IFC label, the second is the
identifier, the third and fourth are the lower and higher points of
the bounding box (resp.), and the fifth depends on the file (in the
duplex file it is the centroid point of the box, in the architecture
model of the office is 'arq', and in the structure model of the office
is 'str'). Several queries, for both models, are available at
\myurl{duplex.pl} and \myurl{office.pl}.  Let us comment a few of them:

\begin{example}[Duplex]
  Fig.~\ref{fig:x3d_a} shows the whole model of the Duplex
  (\myurl{duplex.pl}).  The doors are in green and the rest of the
  objects are in blue (query $Q1$).
  Fig.~\ref{fig:x3d_b} shows the results of the query $Q2$ which
  imposes the constraints \code{Ya#<-4} to select certain doors, and
  \code{Y#>=-7, Y#<-4} to ``create'' a space (unbounded in the axis
  $x$ and $z$) that defines a \emph{slice} of the model.
  Constraints can be used in s(CASP) to reason about unbounded spaces,
  and finer constraints, such as \code{Ya#<-4.002}, can be used
  without performance impact. That is in general not the case with
  other ASP systems.  
\end{example}

\begin{example}[Office]
  \label{exa:query}
  Fig.~\ref{fig:x3d_d} shows the results of the query $Q2$ in
  \myurl{office.pl}, which selects objects of type \emph{IfcBeam} in
  the Architecture model that are not covered by objects in the
  Structural BIM model.  Fig.~\ref{fig:x3d_c} shows those objects that
  intersect the beam (query $Q1$).  If that is the case, the uncovered
  parts are drawn in red.  Note that these parts can be as thin as
  necessary, without negatively impacting performance.
\end{example}

The current development of the BIM reasoner is a proof of concept in
which no optimization techniques have been applied.
Nevertheless, the results from a performance point of view are also
satisfactory. The query in Example~\ref{exa:query} found the first
beam with uncovered parts in \textbf{0.104 sec.} and evaluates the whole
office, by selecting 691 beams from a total of 3639 objects in the
architecture model and detecting the 511 beams not covered by the more
than 1300 objects in the structure model, in 48
  sec.\footnote{\cite{li2020non} reports that ASP4BIM pre-processing
  of 5415 BIM objects takes 99 sec.}

\redsection
\section{Related Work}
\label{sec:related-work}

Many logic-based proposals have been developed to overcome the
limitations of the IFC format and automated tools based on IFC, such
as Solibri Model Checker (SMC) and the Corenet BIM e-Submission by 
\cite{Corenet} to be adapted to different regulations. These
limitations have been attacked using different approaches:
\begin{itemize}
\item Extended query languages to handle IFC data, such as BimSPARQL,
  by \cite{zhang2018bimsparql}, that extends SPARQL with (i) a set of
  functions modeled using the Web Ontology Language (OWL), (ii) a set
  of transformation rules to map functions to IFC data, and (iii) a
  module for geometrical related functions. However, they require to
  pre-process the geometrical information contained in the model and/or
  have limitations to infer new knowledge, e.g., the shortest path
  between two rooms.

\item Minimal proof-of-concept tools, such as the safety checker
  by~\cite{zhang2013a}, the acoustic rule checker by
  \cite{pauwels2011a}, and BIMRL by \cite{solihin2015simplified}, that
  show improved reasoning capabilities of w.r.t.\ commercial
  off-the-shelf BIM Sofware.
  However, all report limitations in the representation of geometrical
  information and/or in the flexibility of the proposal to adapt the
  code and/or the evaluation engine for different scenarios.

\item Translation of building regulation into computer-executable
  formats, such as KBimCode by~\cite{lee2016a} which transcribes the
  Korean Building Act to evaluate building permit requirements. However,
  they report difficulties to translate vague concepts such as
  \emph{accessible routes} (key information such as the function of a
  room is needed to derive the ``accessible routes'').
\end{itemize}
 
To overcome the limitations of these approaches, we propose using a
goal-directed implementation of CASP, because Prolog and bottom-up
implementations of CASP have limitations to modeling vague concepts
and geometrical information simultaneously:

\begin{itemize}
\item \textbf{Prolog:} Since Prolog is based on the least fixed point
  semantics, the different answers generated by independent clauses
  correspond to a single model and are simultaneously true.
  Consider a program containing the facts %
  \code{size_of(r1, small)} and %
  \code{size_of(r1, big)}.
  If \code{size_of/2} is invoked in different parts of the program, it
  may assign two different sizes to the same room, which is against
  the intended interpretation of \code{size_of/2}.
  It is not possible to restrict \code{r1} to have only one of the
  two possible sizes everywhere in the program.  Additional care
  (e.g., explicit parameters) is needed to force this consistency.
  Moreover, it is not easy to make use of default negation in Prolog,
  since its \emph{negation as failure}
  rule 
  has to be restricted to ground calls (e.g., the query in
  Example~\ref{exa:size} is unsound under SLDNF) and it may not
  terminate in the presence of non-stratified negation (e.g., the
  query \code{?- small(r1)} in Example~\ref{exa:bigsmall} does not
  terminate under SLDNF).

  While there exist implementations of Prolog, such as XSB with
  tabled negation,
  that compute logic programs according to the well-founded semantics
  (WFS), the truth value of atoms under WFS can be undefined, e.g.,
  the query \code{?- small(r1)} in Example~\ref{exa:bigsmall} under
  WFS returns \emph{undefined}.

\item \textbf{CASP:} While a goal-directed implementation of ASP
  provides the relevant partial model, standard ASP systems that
  require a grounding phase in the presence of multiple even loops,
  e.g., a unique vague concept referred to various objects, may
  generate a combinatorial explosion in the number of valid stable
  models, reducing the comprehensibility of the results.

  Moreover, these systems  can not (easily)
  handle an unbound and/or dense domain due to the grounding phase.
  Variable domains induced by constraints can be unbound and,
  therefore, infinite (e.g., \code{X#>0} with
  $\mathtt{X} \in \mathds{N}$ or $\mathtt{X} \in \mathds{Q}$).  Even
  if they are bound, they can contain an infinite number of elements
  (e.g. \code{X#>0 $\land$ X#<1} in $\mathds{Q}$ or $\mathds{R}$).
  These problems have been attacked using different techniques:

\begin{itemize}
\item Translation-based methods, such as EZCSP
  by~\cite{balduccini17-ezcsp}, convert both ASP and constraints
  into a theory that is executed in an SMT solver-like manner.
  However, the translation may result in a large propositional
  representation or weak propagation strength.
\item Extensions of ASP systems with constraint propagators, such as
  clingcon by~\cite{banbara17-clingcon3}, and clingo[DL,LP],
  generate and propagate new constraints during the search.
  However, they are restricted to finite domain solvers
  and/or incrementally generate ground models, lifting the upper
  bounds for some parameters.
\end{itemize}

Since the grounding phase causes a loss of communication from
the elimination of variables, the execution methods for CASP systems are
complex.  Explicit hooks sometimes are needed in the language, e.g.,
the \texttt{required} builtin of EZCSP, so that the ASP solver and the
constraint solver can communicate. More details on standard CASP
systems can be found in the paper by~\cite{lierler2021constraint}.

\end{itemize}

Finally, let us analyse a recent proposal for safety analysis,
ASP4BIM by~\cite{li2020non}, which is built on top of \emph{clingo},
a state-of-the-art ASP solver.
ASP4BIM overcomes most of the limitations of
previous BIM logic-based approaches by (i) defining spatial aggregates in
ASP, (ii) maintaining geometries in ASP through a specialised geometry
database extended to support the real arithmetic resolution and
specialised spatial optimisations, and (iii) formalising 3D BIM safety
compliance analysis within ASP.
However, it inherits the limitations of ASP solvers, which require a
grounding phase, when dealing with dense/unbounded domains (needed to
represent time, dimensions, etc.) and/or understanding the answers
due to the number, size, or readability of the resulting models.
While the limitation of dealing with dense domains can be overcome by
using discrete domains, e.g., using integers to represent time-steps
instead of continuous-time, it involves certain drawbacks: as pointed
out by \cite{scasp-iclp2018} this shortcut may impact performance (by
increasing execution run-time of \emph{clingo} by orders of magnitude)
and/or may make a program succeed with wrong answers (due to the
rounding in ASP).

\redsection
\section{Conclusion and Future Work}

We have highlighted the advantages of a well-founded approach to
represent geometrical and non-geometrical information in BIM
models, including specifications, codes, and/or guidelines.
BIM models change during their design, construction, and/or facility
time, by removing, adding, or changing objects and
properties. 

The use of CLP, and more specifically s(CASP), makes it possible to
realize common-sense reasoning by combining geometrical and
non-geometrical information thanks to its ability to perform
non-monotonic reasoning and its support for constructive negation.
Our proposal allows the representation of knowledge involving vague
concepts and/or unknown information and the integration of
\mbox{(non-)geometrical} information in queries and rules used to
reason and define BIM models.

We have identified some future research directions.

\begin{itemize}
\item \textbf{BIM Verification vs BIM Refinement} The design and
  construction of a building is a sequence of decisions
  (setting dimensions, materials, deadlines, etc.)
  each of which reduces the degrees of freedom.
  A model refinement approach would generate a sequence of models
  based on the formal specifications of the regulations, client
  requirements, geometry, etc. Any change in the model chain should be
  consistent upwards,
  keeping the refinement structure.

\item \textbf{Non-Monotonic Model Refinement}
  A monotonic evolution of a BIM model, following model refinement,
  ensures consistency.  The natural flow of architectural development
  requires however the consideration of non-monotonic refinements due
  to unforeseen events, cost overruns, delays, etc.

\item \textbf{Integrating logical reasoning in BIM Software}
  This proposal is an initial step that, together with other proposals
  such as ASP4BIM, may lead to a new paradigm in the refinement of BIM
  models that
  would improve the flexibility and reasoning
  capacity of the current standards.
  Its integration with commercial off-the-shelf BIM Software would
  require efficiency improvement, by adapting s(CASP) execution
  strategy or implementing specialized constraint solvers.

\end{itemize}

\subsection*{Acknowledgements}

We would like to thank Vishal Singh, and Mehmet Yalcinkaya, from
VisuaLynk, and especially Olli Sepp\"anen, who hosted J. Arias at
Aalto University in the summer of 2019, for useful discussions in the
early stages of the work we present in this paper.

\paragraph {\bf Competing interests:} The author(s) declare none.

\vspace{-.5em}

\bibliographystyle{tlplike}

\end{document}